\documentclass[twocolumn]{aastex701}

\begin{document}

\title{DESI as sparse Integral Field Spectrograph – I: Spatially resolved chemical enrichment in star-forming galaxies at $z\leq0.1$}


\author[orcid=0000-0001-5824-1040,gname='Vibhore',sname='Negi']{Vibhore Negi}
\affiliation{Kavli Institute for Astronomy and Astrophysics, Peking University, Beijing, 100871, People's Republic of China}
\email[show]{vibhore.negi18@gmail.com}

\author[orcid=0000-0002-5535-4186,gname='Ravi',sname='Joshi']{Ravi Joshi} 
\affiliation{Indian Institute of Astrophysics (IIA), Koramangala, Bangalore, 560034, India}
\email{rvjoshirv@gmail.com}

\author[orcid=0009-0003-4018-9020,gname='B',sname='Vishnupriya']{B. Vishnupriya} 
\affiliation{Indian Institute of Astrophysics (IIA), Koramangala, Bangalore, 560034, India}
\email{bvishnupriya1707@gmail.com}

\author[orcid=0000-0002-7350-6913,gname='Xue-Bing',sname='Wu']{Xue-Bing Wu} 
\affiliation{Kavli Institute for Astronomy and Astrophysics, Peking University, Beijing, 100871, People's Republic of China}
\affiliation{Department of Astronomy, School of Physics, Peking University, Beijing, 100871, People's Republic of China}
\email{wuxb@pku.edu.cn}

\author[orcid=0000-0002-4176-9145,gname='Hassen',sname='Yesuf']{Hassen M. Yesuf} 
\affiliation{Shanghai Astronomical Observatory, Chinese Academy of Sciences, 80 Nandan Road, Shanghai, 200030, China}
\email{yesufh@shao.ac.cn}

\author[orcid=0000-0001-6947-5846,gname='Luis',sname='Ho']{Luis C. Ho} 
\affiliation{Kavli Institute for Astronomy and Astrophysics, Peking University, Beijing, 100871, People's Republic of China}
\affiliation{Department of Astronomy, School of Physics, Peking University, Beijing, 100871, People's Republic of China}
\email{lho.pku@gmail.com}

\author[orcid=0000-0003-4804-7142, gname='Ayan',sname='Acharyya']{Ayan Acharyya} 
\affiliation{INAF—Astronomical Observatory of Padova, Vicolo dell’Osservatorio 5, IT-35122 Padova, Italy}
\email{ayan.acharyya@inaf.it}

\author[orcid=0009-0005-9999-132X, gname='Ramya',sname='Sethuram']{Ramya Sethuram} 
\affiliation{Indian Institute of Astrophysics (IIA), Koramangala, Bangalore, 560034, India}
\email{ramya@itcc.iiap.res.in}

\author[orcid=0000-0003-2923-1585, gname='Abhijeet',sname='Anand']{Abhijeet Anand } 
\affiliation{Inter-University Centre for Astronomy and Astrophysics, Post  Bag 4, Ganeshkhind, Pune 411 007, India}
\email{abhijeetanand2011@gmail.com}

\author[0000-0001-6676-3842, gname='Michele',sname='Fumagalli']{Michele Fumagalli} 
\affiliation{Universit\'a degli Studi di Milano-Bicocca, Dip. di Fisica G. Occhialini, Piazza della Scienza 3, 20126 Milano, Italy}
\affiliation{INAF - Osservatorio Astronomico di Trieste, via G.B. Tiepolo 11, I-34143 Trieste, Italy}
\email{michele.fumagalli@unimib.it}

\author[0000-0002-4288-599X, gname='Celine',sname='P\'eroux']{Celine P\'eroux} 
\affiliation{European Southern Observatory, Karl-Schwarzschildstrasse 2, D-85748 Garching bei Munchen, Germany}
\affiliation{Aix Marseille Universit\'e, CNRS, LAM (Laboratoire d’Astrophysique de Marseille) UMR 7326, F-13388 Marseille, France}
\email{cperoux@eso.org}

\begin{abstract}

We present a spatially resolved chemical abundance analysis of 2291 star-forming galaxies at $z \leq 0.1$, spanning nearly four orders of magnitude in stellar mass ($8 \le \rm log (M_{\star}/M_{\odot}) \le 11.5$), by exploiting the multi-fibre spectra from the Dark Energy Spectroscopic Instrument (DESI) as a sparse integral field spectrograph. 
In the inner regions ($<2R_e$), the radial gas-phase metallicity profiles show an outward-declining trend for massive galaxies, with the steepest gradient ($\nabla_{log(O/H)}$) $\sim-0.08$ dex/R$_{e}$, whereas low-mass dwarf galaxies exhibit nearly flat profiles ($\nabla_{log(O/H)}\sim-0.02$ dex/R$_{e}$). The large galactocentric radii ($\sim$5 R$_{e}$) probed in this study, reveal flat metallicity profiles near the disk-halo interface. Strikingly, these flat metallicity values are consistent across a wide stellar mass range, likely reflecting the influence of low SFR and metal poor inflows in the outer regions.
The metallicity gradient-- stellar mass relation exhibits a turnover at $\log(M_\star/M_\odot) \sim 10.5$, beyond which gradients become shallower, possibly driven by the chemical equilibrium in the inner disk of massive galaxies and/or dilution from cosmic gas accretion.
At fixed stellar mass, a strong size dependence is observed, where compact galaxies show flatter gradients and higher central enrichment than their extended counterparts.
The abundance gradients are further linked with the stellar age distribution within the galactic disk, where galaxies with younger outskirts show steeper gradients than the ones with older outskirts, consistent with ongoing inside-out disc growth sustaining centrally concentrated chemical enrichment.
These results underscore the interplay of star formation efficiency, stellar feedback, and metal-poor gas accretion in governing the radial chemical structure in galaxies.

\end{abstract}


\section{Introduction}
\label{sec:introduction}
A fundamental aspect of galaxy evolution is the buildup of heavy elements through stellar nucleosynthesis and their dispersal into the interstellar medium via stellar winds and supernovae, progressively enriching the surrounding gas. This enrichment reflects the complex interplay between star formation, gas inflows, and galactic outflows, which collectively regulate the chemical evolution of galaxies across cosmic time \citep{Dave2011,Finlator2008}. The resulting gas-phase metallicity provides a fossil record of a galaxy’s integrated star formation history, as well as the regulatory influence of gas inflows and outflows. Observationally, this chemical record is most commonly accessed through metallicity-sensitive optical emission lines, enabling systematic studies of galaxy populations across cosmic time \citep[and references therein]{Pagel1979,Pagel1992,Pettini2004,kewley2008,Maiolino2011,2020ARA&A..58...99S}.\par

The fiber-based spectroscopic surveys, primarily probing the central regions of galaxies, have revealed several tight scaling relations linking gas-phase metallicity to global galaxy properties. Among the most prominent is the mass–metallicity relation, which demonstrates that more massive galaxies tend to host more metal-rich gas \citep[and references therein]{2004ApJ...613..898T,2019A&ARv..27....3M,Sanchez2019,lara2010fundamental}. Beyond this primary dependence, secondary correlations with star formation rate and gas content point to a more complex interplay between star formation, gas inflows, and feedback processes \citep{2010MNRAS.408.2115M,2013MNRAS.433.1425B}, often interpreted within a self-regulated framework of galaxy evolution \citep{2013ApJ...772..119L}. 
These relations have now been extended over a wide range of redshifts, providing important constraints on the build up of stellar mass and evolution of chemical enrichment, across cosmic time \citep{2006ApJ...644..813E, 2014ApJ...791..130Z}, with recent James Webb Space Telescope (JWST) observations pushing these measurements into the epoch of reionization out to z$\sim$10 \citep{2023ApJS..269...33N, 2024A&A...684A..75C, 2023NatAs...7.1517H}.

The advent of integral field spectrographs (IFS) like Multi Unit Spectroscopic Explorer (MUSE, \citealt{2010SPIE.7735E..08B}), PPKAS IFS Nearby Galaxy Survey (PINGS, \citealt{2010MNRAS.405..735R}), Calar Alto Legacy Integral Field Area (CALIFA, \citealt{2012A&A...538A...8S}), Mapping nearby galaxies at Apache point observatory (MANGA, \citealt{2015ApJ...798....7B}), and Sydney-AAO Multi-object Integral-field spectrograph (SAMI, \citealt{2015MNRAS.447.2857B}), has enabled spatially resolved studies of star formation and feedback processes in galaxies. Several dedicated studies have investigated the internal metallicity gradients and their connection to local physical conditions such as stellar mass surface density and star formation activity 
\citep{PerezMontero2016,barrera2016galaxy,belfiore2017}. It has been found that local disc galaxies, typically exhibit negative metallicity gradients, where the metallicity of a galaxy decreases radially from the centre \citep{1971ApJ...168..327S,VilaCostas1993,Zaritsky1994,Moustakas2010,Rupke2010,Sanchez2012b,Sanchez2014,Ho2015,PerezMontero2016,SanchezMenguiano2016,belfiore2017,SanchezMenguiano2018,Poetrodjojo2018,2020ARA&A..58...99S}. 
Such negative gradients provide strong observational support for the inside-out growth scenario of disc galaxies, in which star formation and chemical enrichment occur earlier and more efficiently in the inner regions, while the outer disk builds up more gradually, remaining relatively metal-poor \citep{1999MNRAS.307..857B,2000MNRAS.313..338P,Sanchez2014}. Recent studies have further corroborated this picture by measuring the ages of localized star-forming regions using ultraviolet to mid-infrared photometry, identifying negative radial gradients in the onset of recent star formation. This indicates that younger stellar populations in the outer disk formed later than those in the central regions, providing direct temporal evidence for inside-out disk assembly \citep{2022MNRAS.515.3270S,2023A&A...673A.147P}.

However, a broader agreement lacks on whether these metallicity gradients correlate with the galaxy parameters like stellar mass, morphology, stellar age, etc., (\citealt[and references therein]{2020ARA&A..58...99S}, \citealt[etc.]{Ho2015,Sanchez2014,SanchezMenguiano2016,PerezMontero2016,belfiore2017}).  \citealt{Sanchez2014,SanchezMenguiano2016,SanchezMenguiano2018} on their analysis of CALIFA and MUSE data for a sample of star-forming galaxies found a characteristic gas phase metallicity gradient that does not correlate with the stellar mass and other galaxy properties. Whereas, studies on the CALIFA, SAMI and MANGA data \citep{PerezMontero2016,Poetrodjojo2018,belfiore2017}, have found a positive (/weak) corelation between the metallicity gradients and stellar mass and morphology type. 
These observations highlight the coupling between the star formation and feedback processes in redistributing the metals within galaxies. Overall, there is no clear consensus on how metallicity gradients relate to stellar mass or morphology, motivating further analysis with larger samples.
Further, studies have reported flattened or even positive metallicity gradients in few low-z galaxies, suggesting that additional processes beyond simple inside-out growth may influence the redistribution of metals within galactic disks \citep{PerezMontero2016}.
Hence, despite major progress from integral field spectroscopic surveys, the physical origin and dominant drivers of metallicity gradients are still not fully understood, and much remains to be explored with new large datasets.

In the present study, we utilize the high multiplexing and spatial coverage of the Dark Energy Spectroscopic Instrument (DESI), which maps multiple star-forming regions within individual galaxies and provides spatially resolved information, effectively serving as a sparsely sampled IFS (see below, Section ~\ref{sec:sample}). Targeting the local galaxies at $z < 0.1$, over around four orders of magnitude in stellar mass ranging from $\rm 8 \le log (M_{*}/M_{\odot}) \le 11.5$, we probe radial variations in metallicity and nitrogen-to-oxygen abundance across a large, statistically representative sample of star-forming galaxies and study their internal chemical structure and evolution. \par

The paper is structured as follows: Section~\ref{sec:data} describes the data and catalogs used in this work and explains the sample selection criteria. The detailed methodology to measure the gas-phase metallicity and chemical abundance gradients is described in Section~\ref{sec:methodology}. Section~\ref{sec:results} highlights our key results from the analysis, followed by a discussion on results in Section~\ref{sec:discussion}. We finally summarize and conclude our findings in Section~\ref{sec:conclusions}. 
Throughout this work, we use a $\lambda$CDM cosmology with $\Omega_{M}$ = 0.286, $\Omega_{\lambda}$ = 0.714 and $H_{0}$ = 69.6 $km^{-1} s^{-1} Mpc^{-1}$.

\begin{figure*}
    \centering
    \includegraphics[width=\textwidth]{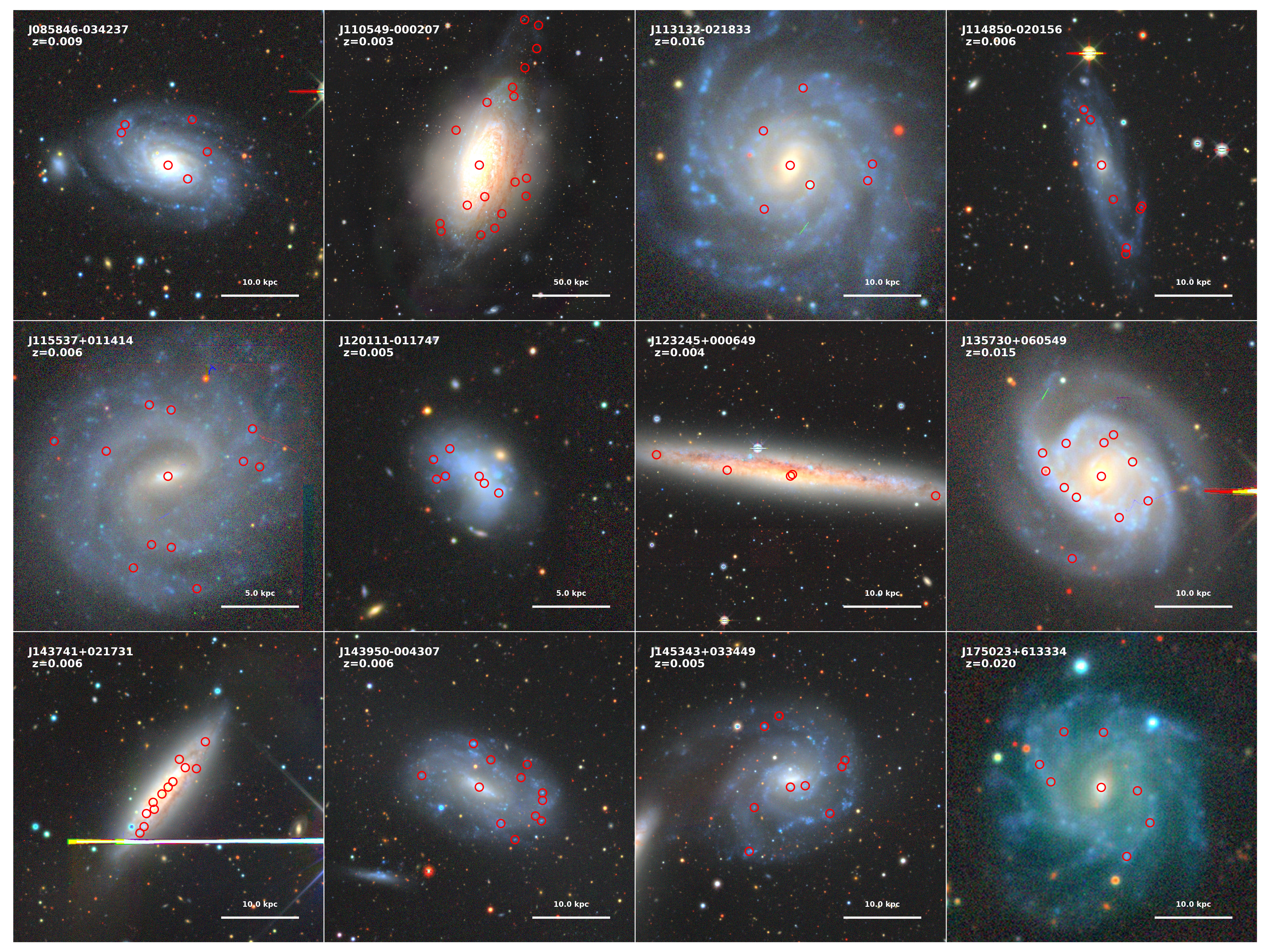}
 \caption{A color composite, $g, r, z$ band  DECaLS image, for few galaxies in our sample, centered on the target galaxy. The red circles indicate the positions of multiple DESI fibers placed on the galaxy. The circle sizes are illustrative and do not correspond to the actual DESI fiber diameter (i.e., 1.5 arcsec).}
         \label{Figure:decals_cutouts_rgb}
\end{figure*}

\begin{figure*}
    \includegraphics[width=\textwidth]{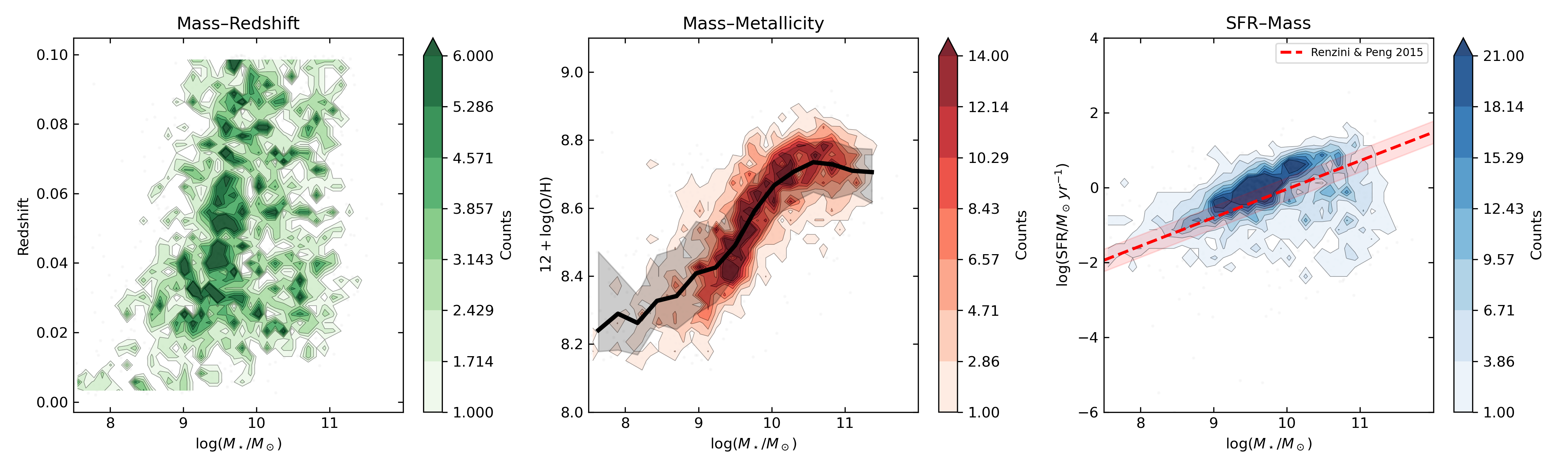}
\caption{Overview of the DESI DR1 star-forming galaxy sample used in this work. The distribution of galaxies is shown with filled contours, colour-coded by number density (counts).
    \textit{Left panel:} Distribution of galaxies in the redshift–stellar mass ($M_\star$) plane. 
    \textit{Middle panel:} Mass–metallicity relation for the sample. The solid black line represents the median gas-phase metallicity in bins of stellar mass, and the shaded region denotes the 16th–84th percentile range at fixed mass.
    \textit{Right panel:} Distribution of galaxies in the star formation rate (SFR)–$M_\star$ plane. The red dashed line shows the star-forming main sequence from \citet{Renzini2015}, with the shaded region indicating $\pm$0.3 dex scatter.}
         \label{Figure:mass_redshift_sample}
\end{figure*}
\section{Data and Sample Selection} \label{sec:data}
\subsection{DESI data}

In this study, we use the first data release (DR1) from the DESI survey, comprising spectra of $\sim$13 million galaxies \citep{DESIDR1}. The DESI is a cosmological survey, conducted over $\sim$14,000 deg$^{2}$ of extra-galactic sky using a multi-object spectrograph mounted on the 4-meter Mayall telescope at Kitt Peak National Observatory. It offers a moderate spectral resolution of $R \sim 2000$--5000, covering a wide wavelength range 3600--9800~\AA\ \citep{2022AJ....164..207D,2023AJ....165....9S,2024AJ....168...95M}. The key physical parameters, including stellar mass, $D_{n}4000$ (an index that traces mean stellar population age), and emission-line measurements for DESI DR1 galaxies with reliable redshifts \citep{AnandRedrock2024}, were obtained from a value-added \textit{Stellar Mass and Emission Line Catalog} (hereafter SMELC)  \citep{2024ApJ...961..173Z}. The stellar masses are derived using \texttt{CIGALE} \citep{Boquien2019}, combining broad-band $g$, $r$, $z$, $W1$, and $W2$ photometry from the DESI Legacy Imaging Surveys (DESI-LS, \citealt{Dey2019}), along with 10 synthetic bands generated from DESI spectra. In brief, the DESI-LS catalogs are constructed using \texttt{The Tractor} \citep{Lang2016}, a forward-modeling framework that fits parametric source models (e.g., PSF, exponential, and de Vaucouleurs profiles) convolved with the local PSF directly to the imaging data. This approach simultaneously constrains source positions, multi-band fluxes, and structural parameters such as the half-light radius and ellipticity, providing consistent model-based measurements \citep{Lang2016,Dey2019}.

To derive the gas phase metallicity, we adopt emission-line fluxes from the DESI value-added \texttt{EmFit} catalog \citep{2025ApJ...982...10P}, which employs non-linear least-squares minimization to simultaneously model prominent nebular emission lines (e.g., H$\beta$, [O\,\textsc{iii}], H$\alpha$, [N\,\textsc{ii}], [S\,\textsc{ii}]) in the continuum subtracted spectra. The stellar continuum is obtained from \texttt{FastSpecFit} \citep{Moustakas2023} which models it using 168 composite stellar population templates of varying age, stellar metallicity, and dust attenuation. The spectra are then corrected for any residual, unmodeled flux by constructing a smooth continuum using a sliding median with iterative outlier-clipping. This approach allows for more flexible line-profile modeling and yields more reliable flux estimates compared to single-component fitting. Furthermore, we measure the [O\,\textsc{ii}]$\lambda\lambda3726,3729$ emission line fluxes, which are not listed in \texttt{EmFit} catalog, by modelling the local continuum with a lower-order polynomial and integrating the line profile over the continuum-subtracted spectrum.
\par

\subsection{Sample selection}
\label{sec:sample}
The DESI instrument, with a large field of view of 8 square degrees, simultaneously captures spectra of 5000 targets using robotically controlled fibers. An important aspect of DESI for this work is its targeting strategy for spectroscopic follow-up. In brief, photometrically identified objects are targeted from the DESI-LS catalog, which is obtained through profile modelling of deep DECaLS images, enabling the separation of smooth structure (bulge, disk) from clumpy star-forming features within a galaxy. This enables spectroscopic measurements at different positions across galaxies, effectively providing spatially resolved information without traditional integral field spectroscopy.

Taking advantage of the unprecedented multiplexing provided by DESI to investigate the spatially resolved characteristics of galaxies, we focused on low-redshift ($z < 0.1$) systems from the SMELC catalogue. Owing to the modest redshift, it offers a greater likelihood of finding several star-forming regions inside the galaxy. For each photometric system, we search for neighboring objects within a projected separation of 100~kpc, and a line-of-sight velocity difference of 500~km~s$^{-1}$. We then identify candidate systems with at least three independent spectroscopic fibers detected within these spatial and redshift tolerances, and with available measurements in the SMELC catalog. This initial selection is intended to isolate systems where multiple fibers probe different regions of the same galaxy.

However, this selection is susceptible to contamination from physically distinct galaxy groups, chance projections of nearby galaxies at similar redshifts, and uncertainties in redshift measurements. 
To mitigate these effects, we visually inspected DESI-LS imaging cutouts for all candidate systems to reject such contaminations, and also demanded one fiber to be positioned at the galaxy's center.
This resulted in a sample of 2880 galaxies, with stellar masses, S\'ersic indices, and half-light radii available in the SMELC catalog, with typical redshifts $0.002 \leq z \leq 0.099$, and stellar mass $5.5 \leq M_{\odot} \leq 11.7$. Among them, only 271 belong to the dwarf regime ($\rm log (M_*/M_{\odot}) < {\mathrm{9.5}}$), primarily due to the comparatively small physical size of dwarf galaxies. To improve the sampling of dwarf galaxies, we broadened the criterion to include galaxies with at least two fibers: a central fiber and an additional fiber in the outer disk. This augments 865 dwarf galaxies, which increases the sample size to 3745 galaxies.
\par

\begin{figure}
    \includegraphics[width=\columnwidth]{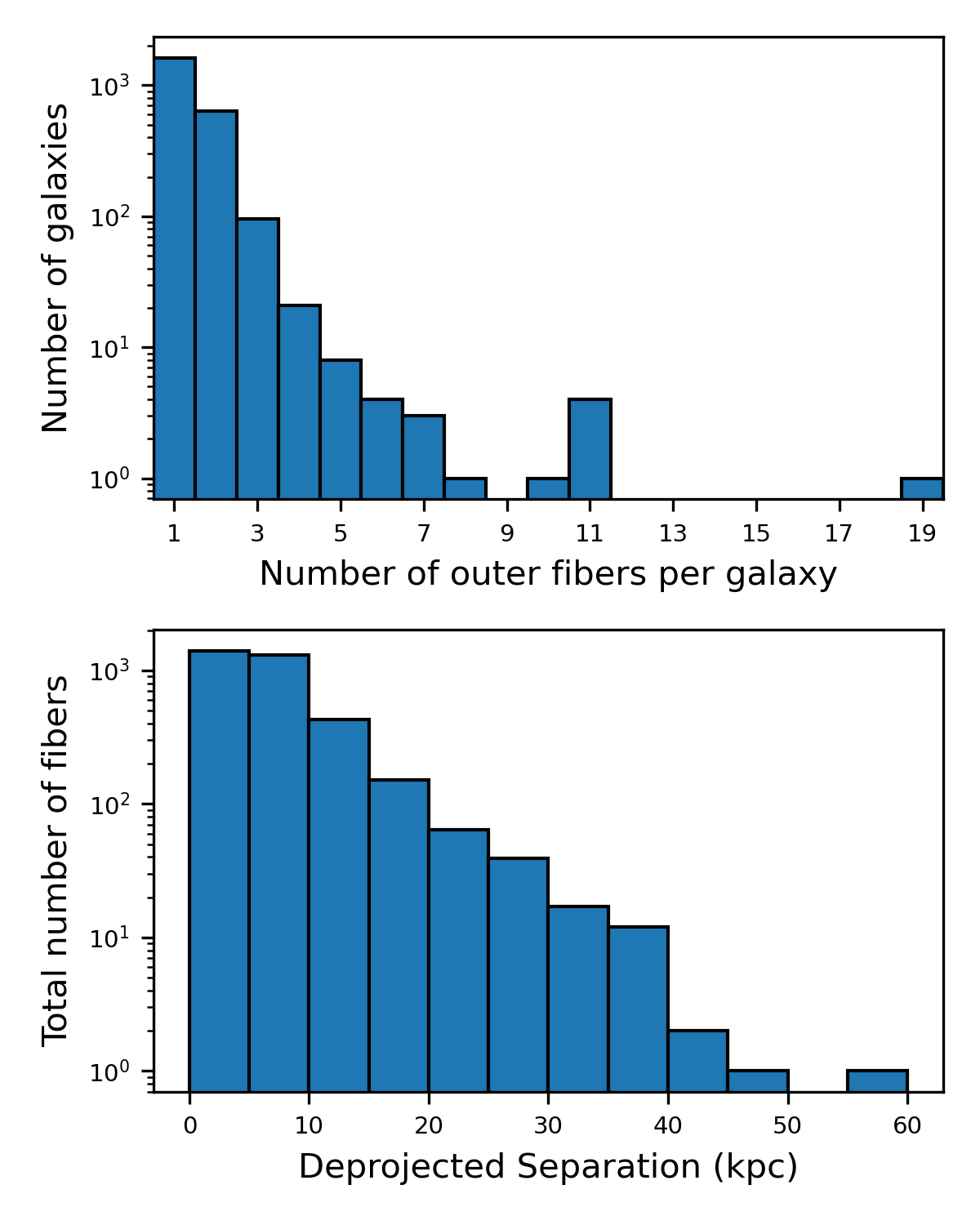}
\caption{ 
    \textit{Top panel:}  Multiplicity distribution of outer fibers per galaxy within the adopted matching criteria ($\leq$100 kpc and $\Delta v \leq 500$ km s$^{-1}$), and after visual inspection. 
    \textit{Bottom panel:} 
    Distribution of deprojected radial separations of outer fiber positions from the galaxy center, measured in bins of 5 kpc.
    }
         \label{Figure:number_distance_sample}
\end{figure}

Finally, to ensure the robustness of the sample and mitigate potential biases, we first exclude the AGNs by performing a BPT classification using the [O\,\textsc{iii}]$\lambda5007$/H$\beta$ versus [N\,\textsc{ii}]$\lambda6583$/H$\alpha$ diagnostic diagram, adopting the demarcation curves of \citet{Kewley2001} and \citet{Kauffmann2003}. 
Next, to ensure reliable metallicity gradients measurements, we remove highly inclined galaxies with inclination angle $> 75^{\circ}$, where projection effects are most severe. The inclination angle is estimated from the observed axis ratio (in DESI-LS) using the standard Hubble formula \citep{Hubble1926}, assuming an intrinsic disk thickness of \(q_0=0.2\). 
We further restrict the sample to star-forming and intermediate systems using the criteria $(\mathrm{D}_{n}4000 < 1.6)$ or an $\mathrm{H}\alpha$ equivalent width (EW) $> 3~\text{\AA}$. 
For reliable metallicity measurements, we additionally require an emission line detection significance of $\ge 2\sigma$. This resulted in a final sample of 2291 star-forming galaxies. Figure~\ref{Figure:decals_cutouts_rgb} shows a representative sample of galaxies from our 
catalog, with red circles marking the positions of DESI fibers placed on each 
system. 
\par


\section{Methodology} \label{sec:methodology}
\subsection{Dust and Extinction correction}
In order to compute the chemical abundances, we first correct the emission-line fluxes for dust attenuation using the Balmer decrement. We estimate the nebular color excess, $E(B-V)$, from the observed H$\alpha$/H$\beta$ ratio assuming an intrinsic Case~B recombination value of $(\mathrm{H}\alpha/\mathrm{H}\beta)_{\rm int}=2.86$ \citep{Osterbrock1989}. The wavelength-dependent attenuation is then evaluated using the extinction curve of \citet{Cardelli1989} with $R_V=3.1$, and the observed line fluxes are de-reddened accordingly before applying the abundance diagnostics. Although line ratios involving nearby transitions are only weakly affected by dust, we apply the extinction correction uniformly to all relevant emission lines for consistency.
\subsection{Measuring gas phase metallicity}

Oxygen, being the most abundant heavy element in the interstellar medium, serves as the primary tracer of gas-phase metallicity in galaxies. Hereafter, we use the terms oxygen abundance and gas-phase metallicity interchangeably. Over the past decades, several calibrations have been developed to estimate oxygen abundance in star-forming galaxies. The most direct method involves measuring the ratio of the weak auroral line [O\,\textsc{iii}]$\lambda4363$ to [O\,\textsc{iii}]$\lambda5007$ \citep{Peimbert1969,Stasinka1978,Pagel1992,Vilchez1996,2006A&A...448..955I,kewley2008}, which provides an estimate of the electron temperature ($T_e$) of the gas. However, the intrinsically weak nature of the auroral line makes it difficult to detect in DESI spectra at typical signal-to-noise ratios, rendering direct $T_e$ measurements impractical for large samples. \par

To overcome this limitation, empirical strong-line calibrations have emerged as robust alternatives, relating strong emission-line ratios to metallicity by anchoring them to $T_e$-based measurements in \texttt{HII} regions. These methods were first introduced by \citet{Alloin1979,Pagel1979}, and have since been extensively developed using empirically calibrated (i.e., $T_e$-anchored) strong-line methods \citep[e.g.,][]{Pilyugin2000,Denicolo2002,Pettini2004,Pilyugin2005,2010ApJ...720.1738P,Marino2013} as well as photoionisation models \citep[e.g.,][]{McGaugh1991,Kewley2002,Dopita2006,Dopita2013,Dopita2016}. 
\par

In this work, we adopt the calibration from \citet[][hereafter M13]{Marino2013}, based on the widely used O3N2 index (first introduced by \citealt{Alloin1979}). The O3N2 index is defined as:
\begin{equation}
\mathrm{O3N2} = \log \left( \frac{[\mathrm{O\,III}]\,\lambda5007 / \mathrm{H}\beta}{[\mathrm{N\,II}]\,\lambda6584 / \mathrm{H}\alpha} \right),
\end{equation}
The corresponding oxygen abundance is given by:
\begin{equation}
12 + \log(\mathrm{O/H}) = 8.533 - 0.214 \times \mathrm{O3N2},
\end{equation}
which is valid over the range $-1 \lesssim \mathrm{O3N2} \lesssim 1.7$ and has an intrinsic scatter of $\sim 0.08$ dex. 
Compared to the original O3N2 calibration by \citet{Pettini2004}, the M13 calibration was derived using a larger sample of H\,{\sc ii} regions with updated direct-method abundances, and provides a revised metallicity scale, particularly relevant at high metallicities \citep{Poetrodjojo2021}. Figure~\ref{Figure:mass_redshift_sample} shows the mass-redshift plane, exhibiting sample completeness down to log($M_{\star}/M_{\odot}) \sim 9$ and $z \sim 0.1$ (left panel), along with the mass-metallicity relation  (middle panel) and the star-formation main sequence (right panel) for the galaxies in our sample.

For a complete understanding of the chemical evolution of galaxies, we combine the metallicity (12+log(O/H)) with the nitrogen abundance (log(N/O)). Unlike oxygen, which is primarily produced in massive stars and released on short timescales, nitrogen has both primary and secondary production channels, with a significant contribution from intermediate-mass stars on longer timescales \citep{2000ApJ...541..660H,VilaCostas1993}. As a result, the (log(N/O)) provides a sensitive tracer for star formation history, gas inflows and outflows, and chemical enrichment timescales of galaxies. 
For log(N/O), we adopt the ON calibration of \citet[hereafter P10]{2010ApJ...720.1738P}, which utilizes strong emission-line fluxes of [O\,\textsc{ii}]$\lambda\lambda3726,3729$, [O\,\textsc{iii}]$\lambda\lambda4959,5007$, and [N\,\textsc{ii}]$\lambda6584$. The relevant line ratios are defined as:
\begin{equation}
R2 = \frac{[\mathrm{O\,II}]\,\lambda\lambda3726,3729}{\mathrm{H}\beta},
\end{equation}
\begin{equation}
R3 = \frac{[\mathrm{O\,III}]\,\lambda4959 + [\mathrm{O\,III}]\,\lambda5007}{\mathrm{H}\beta},
\end{equation}
\begin{equation}
N2 = \frac{[\mathrm{N\,II}]\,\lambda6548 + [\mathrm{N\,II}]\,\lambda6584}{\mathrm{H}\beta},
\end{equation}
\begin{equation}
S2 = \frac{[\mathrm{S\,II}]\,\lambda6717 + [\mathrm{S\,II}]\,\lambda6731}{\mathrm{H}\beta}.
\end{equation}
Following P10, galaxies are classified into cool ($\log N_2 \geq -0.1$), warm ($\log N_2 < -0.1$, $\log(N_2/S_2) \geq -0.25$), and hot ($\log N_2 < -0.1$, $\log(N_2/S_2) < -0.25$) regimes based on these line ratios, and regime-dependent relations are used to estimate $\log(\mathrm{N/H})$ and $\log(\mathrm{O/H})$.
The nitrogen-to-oxygen ratio is then computed as:
\begin{equation}
\log(\mathrm{N/O}) = \log(\mathrm{N/H}) - \log(\mathrm{O/H}).
\end{equation}

Uncertainties in metallicity and nitrogen abundance ratios are estimated using a Monte Carlo approach that propagates measurement errors in the emission-line fluxes. For each galaxy, we generate 1000 realizations of the observed line fluxes by perturbing them according to their associated Gaussian uncertainties, and recompute the derived quantities for each realization (e.g., \citealt{Poetrodjojo2021}).  
The final uncertainties are estimated from the 16th and 84th percentiles of the resulting distributions, corresponding to a $1\sigma$ confidence interval. 

\begin{figure*}[!htbp]
    \includegraphics[width=\textwidth]{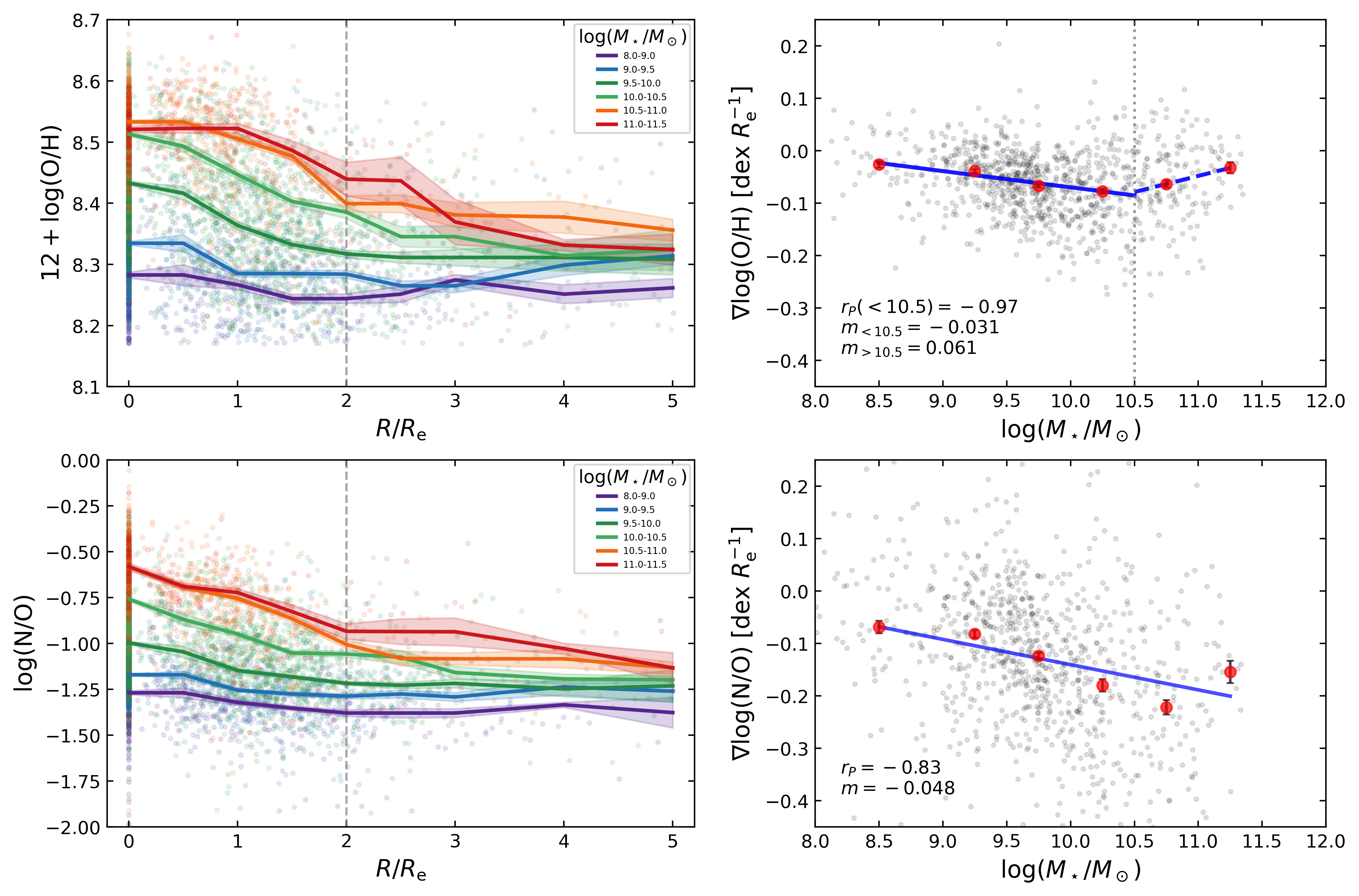}
 \caption{{\it Top Left panel:} Radial gas-phase metallicity profiles, 
$12+\log(\mathrm{O/H})$ (using the M13 calibration based on O3N2), as a function of normalized radius $R/R_{\mathrm{e}}$, in stellar mass bins. 
Solid lines show the median metallicity in radial bins, with shaded regions indicating the uncertainty on the median (including the scatter of points, measured via the MAD, and Monte Carlo errors from individual measurements). Faint points represent individual fiber measurements.
{\it Top Right panel}: Metallicity gradients (dex $R_{\mathrm{e}}^{-1}$) as a function of stellar mass. Grey points show individual galaxies gradients (calculated within $0<R_{e}\leq2.0$), red circles denote median gradients in 0.5 dex stellar mass bins with their associated uncertainties, and the dashed line shows a linear fit to the binned medians for log(M$_\star$/M$\odot$ $<$10.5). The Pearson correlation coefficient ($r_{p}$) and the slope ($m$) to the gradient - mass fit is written in lower left.
{\it Bottom Left panel:} Same as in Top left but for nitrogen abundance profile, calculated using P10 calibrations.
{\it Bottom right panel:} Same as in Top right but for nitrogen abundance gradients.}
         \label{Figure:metallicity_gradient_plot}
\end{figure*}
\subsection{Measuring radial gradients}
To quantify radial variations in chemical abundances, we first deproject the angular separation of each fiber into the disk plane of the galaxy by correcting for its position angle and inclination. The resulting deprojected radius $R$, is normalized by the galaxy's effective radius $R_e$, drawn from the DESI-LS imaging catalog, yielding $R/R_e$.
This normalization allows consistent comparison across galaxies of different physical sizes. We then construct normalized radial profiles of both O/H and N/O using $R/R_e$ at different fiber positions across galaxies. 
A distribution of the number of outer fibers per galaxy, and their deprojected distances (in kpc) from the galaxy center, is shown in top and bottom panels of Figure~\ref{Figure:number_distance_sample}, respectively.

To explore the dependence on stellar mass, galaxies are divided into six mass bins of $\log(M_\star/M_\odot)$ = $(8.0\text{--}9.0)$, $(9.0\text{--}9.5)$, $(9.5\text{--}10.0)$, $(10.0\text{--}10.5)$, $(10.5\text{--}11.0)$, and $(11.0\text{--}11.5)$, comprising 327, 492, 611, 429, 332, 100, galaxies, respectively. Radial profiles are constructed independently within each mass subset by binning the data in normalized radius using fixed intervals of $[0, 0.5, 1.0, 1.5, 2.0, 2.5, 3.0, 4.0, 5.0]R_{\mathrm{e}}$. Within each radial bin, we compute the median metallicity (or log(N/O)) to reduce the impact of outliers and measurement noise. The central galaxy is treated separately and included explicitly as the innermost point, such that each profile consists of the central value followed by median values in successive radial bins. Interestingly, the high-multiplexing  in DESI enabled us, for the first time, to map the metallicity profiles out to $\sim 5\,R_{\mathrm{e}}$. To mitigate the impact of limited fiber statistics beyond $\sim 3Re$, we additionally define three coarse stellar mass bins of $(8.0\text{--}9.5)$, $(9.5\text{--}10.5)$, and $(10.5\text{--}11.5)$. 
The median metallicity in each mass bin is computed, with uncertainty estimated using the median absolute deviation (MAD).\par


Next, we estimated the radial metallicity gradient for each galaxy by anchoring the central metallicity at $r = 0$ and computing the median metallicity of outskirts in bins of projected separation normalized by $R_{\rm e}$. 
Given the limited filed-of-view of IFUs, the radial metallicity gradients in the literature are largely measured within $R/R_{\mathrm{e}} \leq 2 -3$, covering $\sim$ 85 per cent of the galaxy's light for a pure exponential (n=1) system \citep{Graham2005}.
Therefore, to compare with literature measurements, we restricted the fit to $R/R_{\mathrm{e}} \leq 2.0$. To ensure uniform radial coverage and minimize the possible biases due to sparse fiber sampling, we demand at least two radial points mapping inner and outer region of galaxy at a minimum separation of $\sim 1~R_{\mathrm{e}}$, for measuring the gradients. A linear least-squares fit to this profile gives the gradient slope in units of dex\,$R_{\rm e}^{-1}$, per galaxy.




\section{Results\label{subsec:results}}
\label{sec:results}
\subsection{Radial metallicity profiles and gradients}
\label{sec:radial_profile_gradients}
\begin{figure}
    \includegraphics[width=0.45\textwidth]{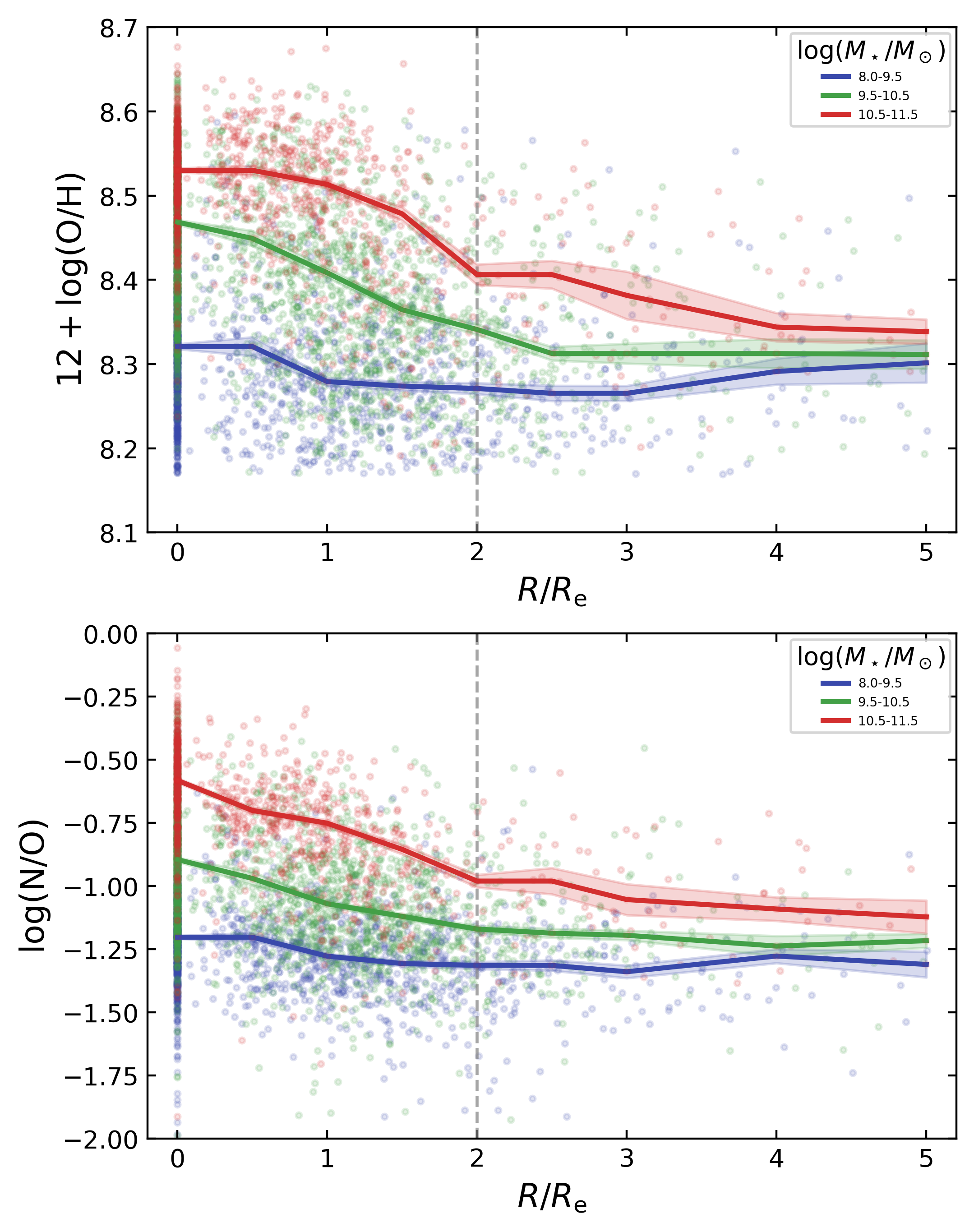}
 \caption{The radial gas-phase metallicity profiles (top panel), 
and the nitrogen abundance profile (bottom panel), as a function of normalized radius $R/R_{\mathrm{e}}$, in coarse stellar mass bins. Solid lines show the median metallicity in radial bins, and shaded regions indicate the uncertainty on the median including the scatter of points, measured via the MAD, and Monte Carlo errors from individual measurements). Faint points represent individual measurements.}
         \label{Figure:metallicity_gradient_plot_coarse}
\end{figure}

Using the calibrations and methodology described in Section~\ref{sec:methodology}, we derive the radial profiles of oxygen and nitrogen abundances for the 2291 galaxies in our sample. The resulting profiles are shown in the top-left and bottom-left panels of Figure~\ref{Figure:metallicity_gradient_plot} for $12 + \log(\mathrm{O/H})$ and $\log(\mathrm{N/O})$, respectively. Individual measurements are plotted as scatter points, colour-coded by their corresponding stellar mass bins. The median trends for the mass bins are represented by solid lines, and extend out to $5R_e$. In addition, the radial profiles for the oxygen and nitrogen abundances in the coarse mass bins are shown in the top and bottom panels of Figure~\ref{Figure:metallicity_gradient_plot_coarse} respectively.

At fixed radius, high mass galaxies exhibit systematically higher values of $12 + \text{log(O/H)}$ and $\text{log(N/O)}$, and this offset persists over a radial range of $\sim 3R_{e}$.
Both oxygen and nitrogen abundances exhibit negative radial gradients across all stellar mass bins, decreasing systematically with increasing radius. An exception exists at low-mass galaxies ($\log(M_\star/M_\odot) < 9.0$, and also $\log(M_\star/M_\odot) < 9.5$), which have radial profiles that are almost flat 
throughout the observable disc. 
The gradients are most pronounced within the inner $\sim2R_{e}$, where the profiles show a smooth and monotonic decline. The radial trends appear comparatively flatter and seem to overlap at outer radii beyond  $\sim 3 R_{e}$, within $2\sigma$ level. This argument is further supported by the profiles in coarse mass bins where more data points contribute to the median trends in each radial bin (see Figure~\ref{Figure:metallicity_gradient_plot_coarse}). 

The metallicity gradient - stellar mass relations based on $12 + \log(\mathrm{O/H})$ (top-right) and $\log(\mathrm{N/O})$ (bottom-right) for individual galaxy (grey points) and median gradients (red points) are shown in Figure~\ref{Figure:metallicity_gradient_plot}, respectively.
%
%
The  (O/H) follows a negative gradient over the stellar mass range of $\log(M_\star/M_\odot) \sim 8.0$–10.5 (with $\nabla \text{log(O/H)}$ going from $\sim0.02~\text{dex}/R_{e}$ to $\sim0.08~\text{dex}/R_{e}$). It shows a clear break around $\log(M_\star/M_\odot) \sim 10.5$, with progressively shallower gradients with increasing stellar mass ( i.e., $\nabla \text{log(O/H)}\sim0.06~\text{dex}/R_{e}$ and $\sim0.03~\text{dex}/R_{e}$, for the two highest mass bins). 
%
%
%
A weaker global Pearson correlation coefficient ($r_{P}\sim -0.33$, with $p_{value}=0.51$) and a strong correlation of ($r_{P}\sim -0.97$ with $p_{value}=0.02$) for $\log(M_\star/M_\odot)\leq10.5$, for the O/H gradients, reflects this departure from a single linear trend at high stellar masses.
This is further evident from the gray scatter points in the background which show a reversal toward less negative gradients in the highest-mass systems.
In contrast, the $\log(\mathrm{N/O})$ gradients steepen monotonically across most of the stellar mass range ($\nabla \text{log(O/H)}$ going from $\sim0.02~\text{dex}/R_{e}$ to $\sim0.08~\text{dex}/R_{e}$), with a Pearson correlation coefficient of $r_{P} = -0.83$ (with $p_{value}=0.04$) and a slope of $m = -0.048$ dex $R_e^{-1}$ per dex in mass. Unlike the oxygen abundance gradients, no clear trend reversal is observed; except a single point for the highest mass bin that suffers from low-number statistics, thus 
should be interpreted with care.


\subsection{Correlation between nitrogen and oxygen abundances}
\label{sec:nitrogen_vs_oxygen_abundance}
\begin{figure}
 \includegraphics[width=\columnwidth,keepaspectratio]{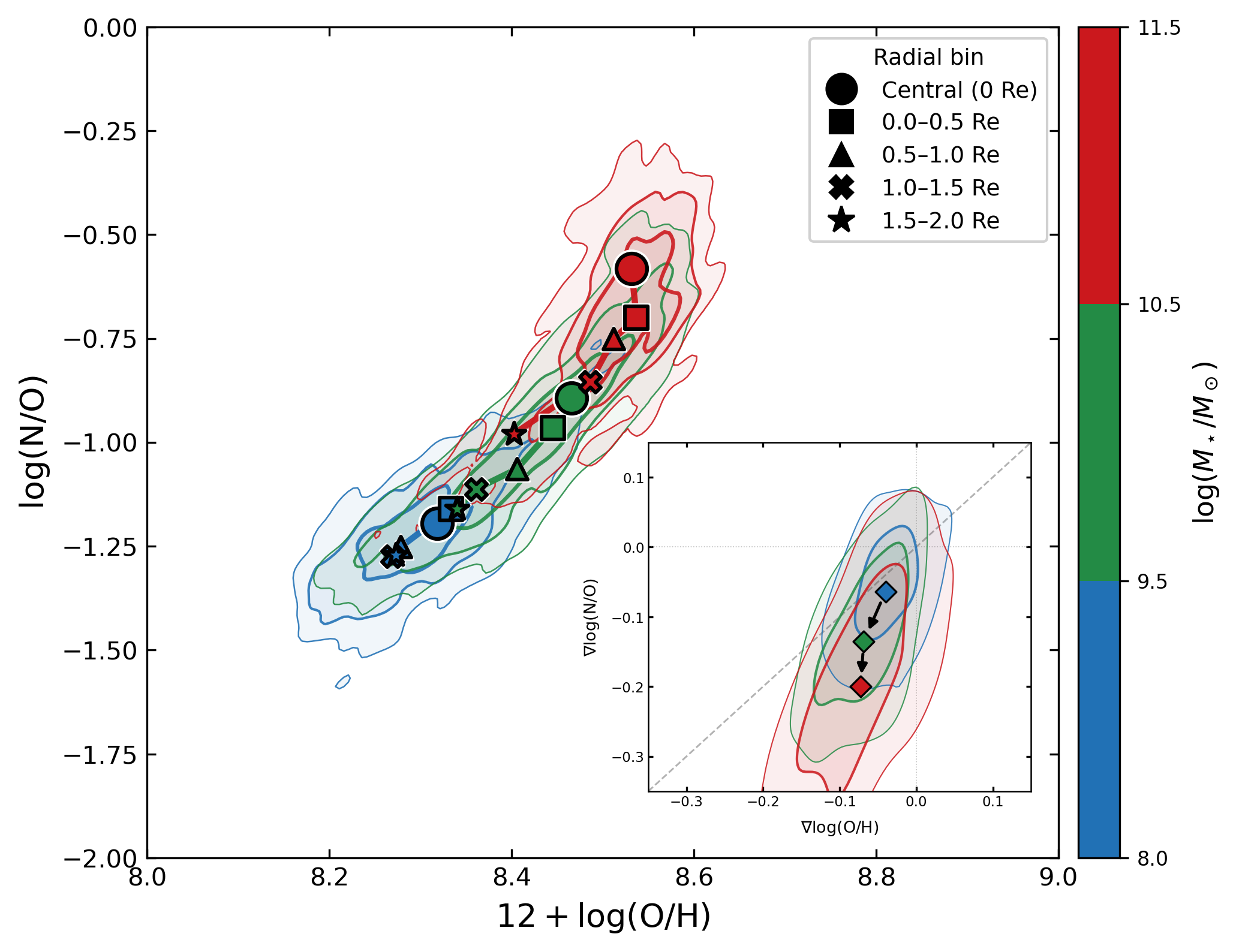}
 \caption{The $\log(\mathrm{N/O})$ versus $12+\log(\mathrm{O/H})$ plane for our sample split by stellar mass. KDE contours show the density distribution of all measurements within $2R_e$. Radial tracks connect median abundances from the galaxy centre outward to $2R_e$. The insets show the gradient space ($\nabla\log(\mathrm{N/O})$ vs.\ $\nabla\log(\mathrm{O/H})$), with arrows connecting median gradients from low to high bins.}
         \label{Figure:metallicity_NO_plot}
\end{figure}

Figure~\ref{Figure:metallicity_NO_plot} shows the $\log(\mathrm{N/O})$ versus $12+\log(\mathrm{O/H})$ plane for our sample split by stellar mass. KDE contours show the density distribution of all individual measurements within $2R_e$, with three coarse mass bins shown in different colors. A clear positive correlation is observed between $12+\log(\mathrm{O/H})$ and $\log(\mathrm{N/O})$, with higher-mass galaxies occupying systematically higher N/O and O/H. Radial tracks progress coherently from the central regions (circle; higher O/H, higher N/O) toward the outer regions (star symbol; lower O/H, lower N/O), exemplifying the negative metallicity gradients reported above. 
Notably, the slope of the N/O--O/H sequence steepens toward higher stellar masses, reflecting the increasing contribution of secondary nitrogen nucleosynthesis at higher oxygen abundances — where nitrogen production becomes proportional to the existing metal content, driving N/O to rise more steeply with O/H. At lower masses and metallicities, the N/O is dominated by primary nitrogen production, and hence has less dependence on O/H, than at high metallicity.
The insets show the corresponding gradient space 
($\nabla\log(\mathrm{N/O})$ vs.\ $\nabla\log(\mathrm{O/H})$), colour-coded by the same mass bins. At low stellar masses and metallicities (blue), the median gradients of $\log(\mathrm{N/O})$ and $\log(\mathrm{O/H})$ lie close to the 1:1 line, indicating that the two gradients are of comparable magnitude. At high stellar masses and metallicities (red), the $\nabla\log(\mathrm{N/O})$ gradient becomes significantly steeper than $\nabla\log(\mathrm{O/H})$, consistent with the enhanced role of secondary nitrogen nucleosynthesis in the central, metal-rich regions of massive galaxies, where N/O rises more rapidly towards the center than O/H.
\begin{figure}
    \includegraphics[width=\columnwidth,keepaspectratio]{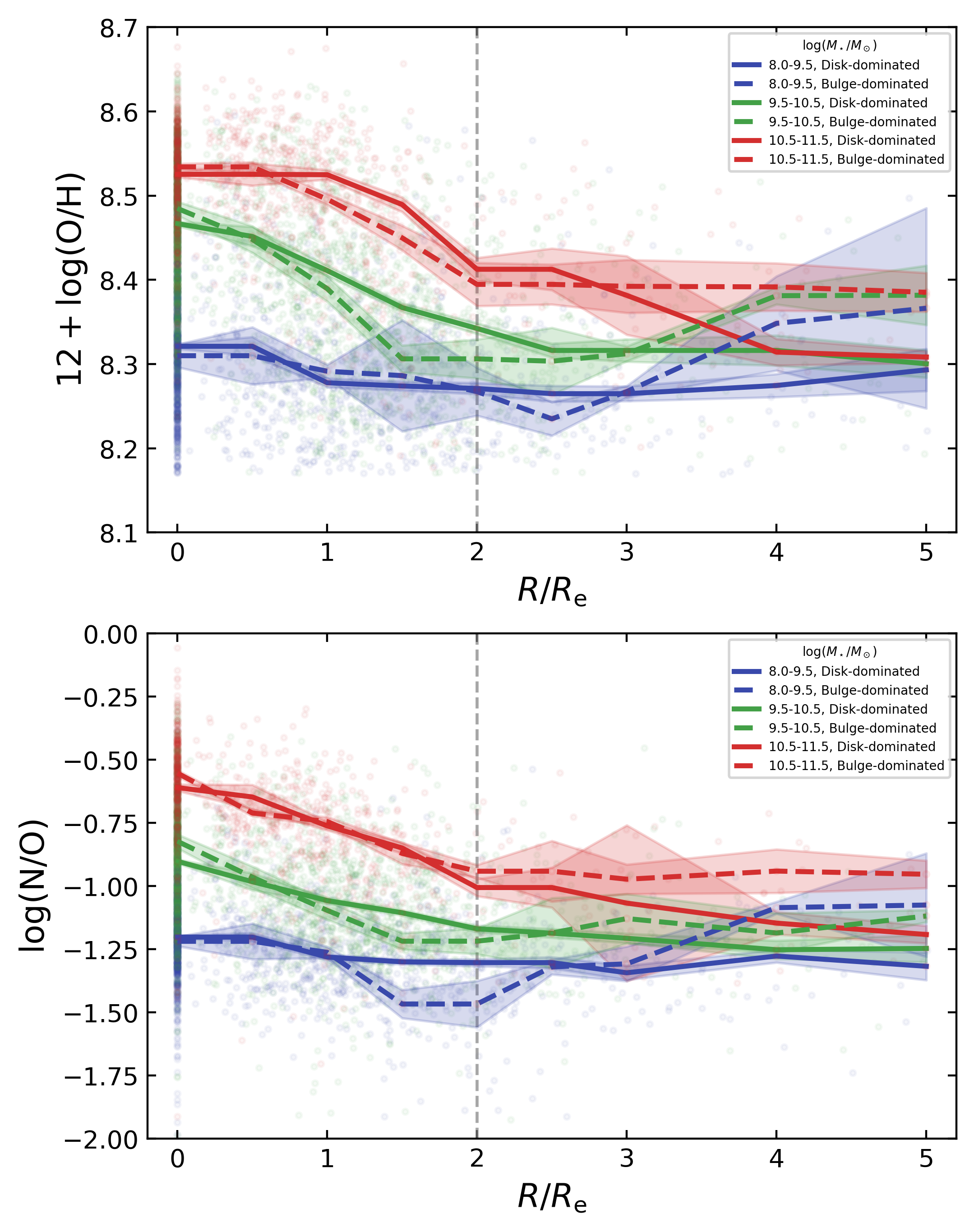}
 \caption{Radial metallicity profiles (top panel), and nitrogen abundance profiles (bottom panel) split by stellar mass and morphology. Solid lines correspond to disk-dominated galaxies (n$<$2.5), while dashed lines correspond to bulge-dominated galaxies (n$>$2.5). Shaded regions indicate the 1$\sigma$ scatter of the binned medians.}
         \label{Figure:metallicity_morphology}
\end{figure}
\subsection{Dependence on galaxy morphology and size}
\label{sec:galaxies_morphology}

To investigate the morphological dependence of the radial metallicity distribution, we further divided the galaxy sample into disk-dominated and bulge-dominated systems using the Sérsic index as a structural proxy. Galaxies with Sérsic index $n < 2.5$ were classified as disk-dominated, while those with $n > 2.5$ were considered bulge-dominated \citep{2003MNRAS.343..978S}. We performed the analysis for the three coarse stellar mass bins defined earlier: $8.0 \leq \log(M_\star/M_{\odot}) < 9.5$, $9.5 \leq \log(M_\star/M_{\odot}) < 10.5$, and $10.5 \leq \log(M_\star/M_{\odot}) < 11.5$.
As evident in Figure~\ref{Figure:metallicity_morphology}, we find no statistically significant difference in the radial O/H and N/O profiles between disk-dominated and bulge-dominated galaxies at fixed stellar mass. Both the central metallicities and radial profiles are consistent within uncertainties across all three mass bins, for both abundance tracers. The mass-dependent trends — higher central metallicity and steeper gradients at higher stellar mass — are recovered consistently in both morphological subsamples, confirming that stellar mass remains the primary driver of the radial chemical structure. We caution however that the bulge-dominated subsample contains significantly fewer galaxies per mass bin than the disk-dominated subsample, which results in larger uncertainties on the bulge profiles and limits our ability to detect any subtle morphological dependence.

\begin{figure}
    \includegraphics[width=\columnwidth,keepaspectratio]{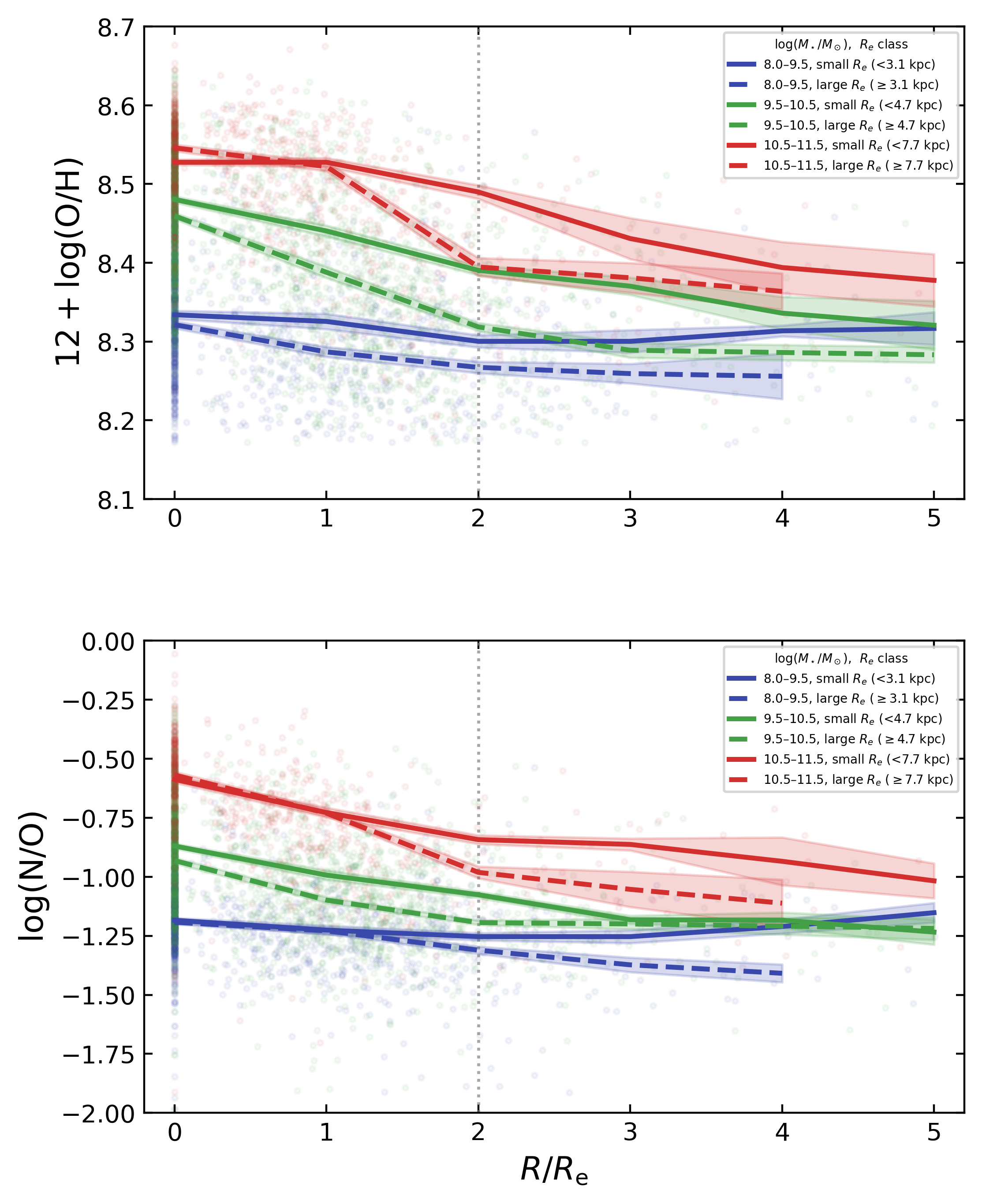}
 \caption{Radial metallicity profiles (top panel), and nitrogen abundance profiles (bottom panel), split by stellar mass and galaxy size. Solid lines correspond to larger galaxies i.e., with radii larger than the median $R_{e}$ of that bin ($R > R_{e,median}$), while dashed lines correspond to smaller galaxies ($R < R_{e,median}$). Shaded regions indicate the 1$\sigma$ scatter of the binned medians.}
         \label{Figure:metallicity_size}
\end{figure}
We further explored the dependence of metallicity on galaxy size by subdividing galaxies in each coarse mass bins into two sub-samples of small and large galaxies - split by the median half light radii of that bin. In each bin, both the sub-samples are matched in the mass and $D_{n}4000$ plane within a tolerance of 0.1 dex in mass and 0.05 in $D_{n}4000$, ensuring that differences in mass and stellar population age do not bias the comparison. A two-sample Kolmogorov--Smirnov tests on the two populations, yielded $D_\mathrm{KS} \leq 0.061$ and 
$p \geq 0.712$ for stellar mass, and $D_\mathrm{KS} \leq 0.042$ and $p \geq 
0.944$ for $D_{n}4000$, across the three mass bins, respectively, 
confirming that the two size sub-samples are statistically indistinguishable in 
their controlled properties. As shown in Figure~\ref{Figure:metallicity_size}, in each mass bin, the smaller (or compact) galaxies are found to have higher metallicities and flatter slopes compared to their larger counterparts, at each given $R/R_{e}$. This size - metallicity offset is observed consistently across both tracers - $12+\log(\mathrm{O/H})$ and $\log(\mathrm{N/O})$ - and across all mass bins. The offset is most pronounced in the intermediate mass bin ($\log\,M_\star/M_\odot = 9.5$--$10.5$), where the separation between the two populations is largest and statistically most robust given the greater number of galaxies in this bin. In the high-mass bin ($\log\,M_\star/M_\odot = 10.5$--$11.5$), a clear offset is also present (beyond $R/R_e \gtrsim 1$)), with compact galaxies maintaining elevated abundances out to large radii ($R/R_e \leq 4$), though the large uncertainty regions reflect the smaller sample size at these masses. In the low-mass bin ($\log\,M_\star/M_\odot = 8.0$--$9.5$), the two size sub-samples are most similar to each other, with only a marginal central metallicity offset and comparatively flat profiles across the full radial range probed. However, the $D_{n}4000$ profiles of both populations in each bin, were found to be broadly similar.
\begin{figure*}
    \includegraphics[width=\textwidth]{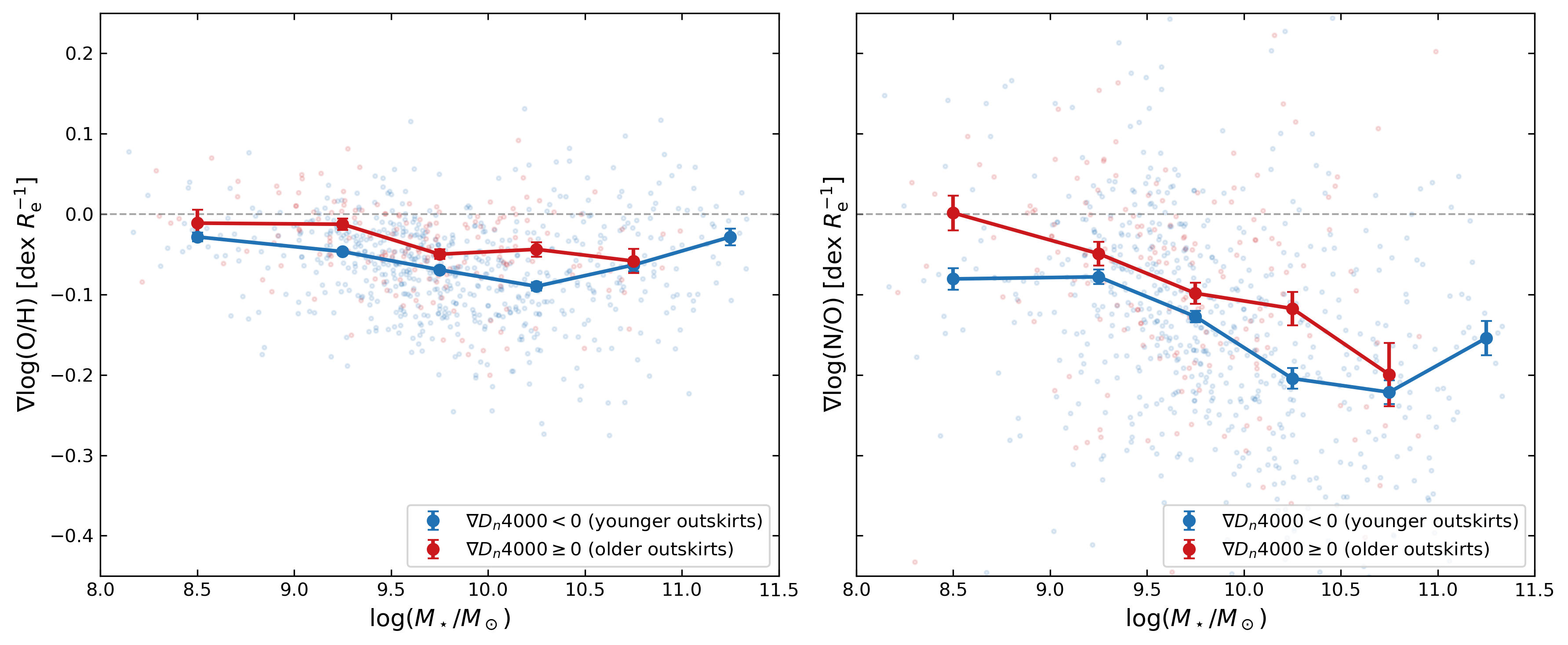}
 \caption{ \textit{Left:} Oxygen abundance gradients as a function of stellar mass for galaxies with negative $\nabla D_{n}4000$ (younger outskirts, blue) and positive $\nabla D_{n}4000$ (older outskirts, red). \textit{Right:} Same as in left panel but for $\log(\mathrm{N/O})$.
  }
         \label{Figure:dn400_split_gradients}
\end{figure*}
\subsection{Dependence on stellar population age}
\label{sec:dn4000_gradient}
To investigate the dependence of the metallicity gradients on the radial 
structure of stellar age, 
we compute the radial gradient of $D_{n}4000$ for each galaxy and divide the sample into two subsets: galaxies with $\nabla D_{n}4000 < 0$ (younger outskirts, consistent with inside-out formation) and galaxies with $\nabla D_{n}4000 \geq 0$ (older outskirts, consistent with more uniform star formation histories). As shown in  Figure~\ref{Figure:dn400_split_gradients}, galaxies with older outskirts exhibit systematically shallower $12+\log(\mathrm{O/H})$ and $\log(\mathrm{N/O})$ gradients across all stellar mass bins. Galaxies with younger outskirts show steeper gradients at all masses, with a clear turnover in $\log(\mathrm{O/H})$ gradients above $\log(M_{\star}/M_{\odot}) \sim 10.5$, consistent with the break identified in the full sample. For $\log(\mathrm{N/O})$ gradients, the younger outskirts population exhibits a more pronounced mass dependence, with gradients falling more steeply across the full mass range compared to the older outskirts population. A tentative flattening at $\log(M_{\star}/M_{\odot}) > 11$ is visible in the younger outskirts population, consistent with the marginal deviation noted in the full sample; however, with relatively large uncertainties.

\begin{figure}
 \includegraphics[width=\columnwidth,keepaspectratio]{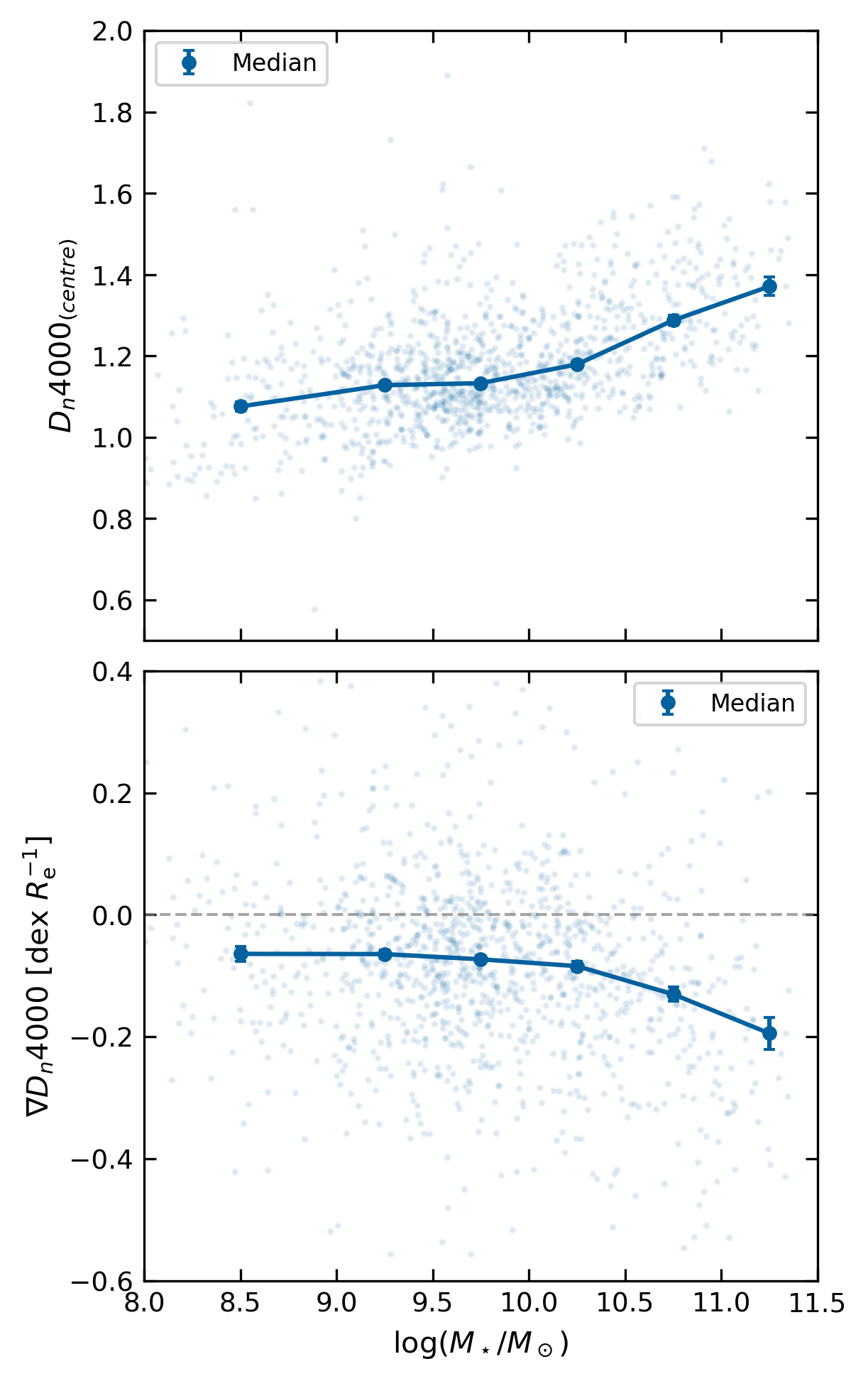}
 \caption{\textit{Top panel:} $D_{n}4000$ as a function of stellar mass. Individual points are shown as blue scatter, and the blue line shows the median connecting median of each bin. \textit{Bottom panel:} Same as in top panel but for $D_{n}4000$ gradients. }
         \label{Figure:dn4000_mass_plot}
\end{figure}
~
\section{Discussion}
\label{sec:discussion}

\subsection{Radial profiles of chemical abundances}

Our results, based on a large sample of 2291 galaxies, confirm the presence of negative gradients in both oxygen and nitrogen abundances for the majority of galaxies, particularly within the inner $\lesssim 2.0\,R_{e}$ region. As shown in Figure~\ref{Figure:metallicity_gradient_plot}, radial metallicity profiles of 819 low-mass dwarf galaxies ($\leq 9.5 M_{\odot}$), exhibit comparatively shallow gradients in $12 + \log(\mathrm{O/H})$, extending out to $\sim 1.5$--$2.0\,R_{e}$. A similar behaviour is also observed in their nitrogen abundance profiles. 
Such flattening of metallicity profiles in low-mass galaxies has been seen commonly in simulations (\citealt{2017MNRAS.466.4780M}, see, however, \citealt{2016MNRAS.456.2982T}), and is likely a consequence of their shallow gravitational potentials, which make stellar feedback and galactic winds particularly effective at driving gas outflows and redistributing metals throughout the galaxy. The radial profiles of local star-forming galaxies in MaNGA and SAMI surveys also show similar flattening in the low mass end \citep[][]{2016MNRAS.458.3466V,Poetrodjojo2018,2020ApJ...890L...3S}.

Towards higher stellar masses, the metallicity profiles (as well as the $\log(\mathrm{N/O})$ profiles) become progressively steeper, as evident in the top-left and bottom-left panels of Figure~\ref{Figure:metallicity_gradient_plot} and the top and bottom panels of Figure~\ref{Figure:metallicity_gradient_plot_coarse}. Such steepening of the metallicity profiles with increasing stellar mass is likely driven by the deeper gravitational potentials of massive galaxies, which promote more efficient retention of metals and enhanced chemical enrichment in their central regions, leading to progressively lower metallicities toward larger galactocentric radii \citep[and references therein]{2019A&ARv..27....3M}.
This trend of increasing gradient steepness with stellar mass has been claimed in previous studies (\citealt{belfiore2017,2020A&A...636A..42M,Poetrodjojo2018,2025A&A...693A.150K}), on the MaNGA and SAMI galaxies, using the R23 and O3N2 calibrations, as well as the robust direct $T_{e}$ method. However, no such significant dependence of the gradients on the stellar mass was found by \citet{Sanchez2014,SanchezMenguiano2016,SanchezMenguiano2018}, using the O3N2 based calibrations, and a marginal detection of the dependence of O/H and N/O gradients on stellar mass, was detected by \citet{PerezMontero2016}. Hence, our results with an enlarged sample confirm the dependence of abundance gradient on the stellar mass of the galaxy. 

Notably, the metallicity profiles of massive galaxies show a flattening in the innermost regions ($\leq 1~R_{e}$), however, the N/O profiles remain steep. Such an inner disc flattening may likely arise due to the consumption of a significant fraction of interstellar gas through earlier episodes of efficient star formation, reaching a chemical equilibrium and slowing the enrichment process \citep{belfiore2017, SanchezMenguiano2018, 2021A&A...655A..58Z}.
%
%
%
%
The persistence of steepness in the N/O profiles is potentially linked to delayed secondary production of nitrogen by intermediate-mass stars, which continues to accumulate on Gyr timescales. This is further supported by Figure~\ref{Figure:dn4000_mass_plot}, which shows that more massive galaxies have both higher central $D_{n}4000$ values and steeper negative $D_{n}4000$ gradients — a direct indication that their centres host older stellar populations relative to their outskirts, 
consistent with earlier and more efficient central star formation and subsequent gas exhaustion.

At large galactocentric radii of $\gtrsim 2 \,R_{\mathrm{e}}$, the abundance profiles start to show a mild flattening, for all mass bins, becoming significantly shallower beyond $\gtrsim 3 \,R_{\mathrm{e}}$. These trends are further supported by the radial profiles derived from the coarse mass bins having substantially more data points per radial bin. An important aspect of our analysis is the extension of the radial profiles out to the disc-halo interface of $\sim5~R_{e}$, a regime that has barely been charted in previous studies. Interestingly, for galaxies across our full mass range, the metallicity at these radii converges around similar values, implying that the outer regions of galaxies are more chemically homogeneous.

Observations of individual galaxies \citep{MartinRoy1992,Vilchez1996,RoyWalsh1997,vanZee1998,Bresolin2009,Werk2010,Werk2011,RosalesOrtega2011,Marino2012,Sanchez2012b,LopezSanchez2015} have also reported such flattening. \citet{Sanchez2014,SanchezMenguiano2016} have also observed such flattening beyond $2R_{e}$ and found it to persist across all star-forming galaxies regardless of morphology, presence of bars, or luminosity, suggesting that it may be a common feature of disk galaxies.

Such flattening is broadly consistent with expectations for the outer disk and disk--halo interface, where processes such as metal-poor gas accretion, radial gas flows, and mixing can act to reduce radial variations in chemical abundances at large radii. \citep[see][and references therein]{2017ASSL..434..145B}. 
Cosmological simulations like TNG-50 \citep{Garcia2023}, and zoom-in FOGGIE simulations \citep{2025ApJ...979..129A}, have also shown such flattening, where it is attributed to the inner-CGM metallicity flooring, formed by regulation of pristine gas infalling in the outer disc balancing the metal rich outflows.  The present study highlights the importance of probing the metallicity enrichment out to $\sim5R_e$ and beyond, to observationally constrain the disk--halo interface, a regime that remains largely inaccessible to conventional observations but is routinely explored in simulations.

\subsection{Correlation of gradients with stellar mass, galaxy morphology and stellar age structure}

We report a strong linear correlation between the metallicity gradient and stellar mass, with a turnover
at $\rm log (M_*/M_{\odot}) \sim 10.5$ (see Section~\ref{sec:results}).
A similar turnover at $\rm log(M_\star/M_{\odot}) > 10 - 10.5$, has also been reported in a handful of studies (see, \citealt{belfiore2017,Poetrodjojo2021,2025A&A...693A.150K,2021ApJ...923...28F}), as well as cosmological simulations \citep{Ibrahim2025}. The observed turnover in the MZGR is primarily driven by the flattening of metallicity profiles in the innermost regions ($\leq 1~R_{e}$) of massive galaxies, where central enrichment has stalled and the gradient loses its sensitivity to stellar mass.
Alternatively, this can be understood within the framework of an analytical model by \citet{2021MNRAS.504...53S,2021MNRAS.502.5935S}, who describe how the dominant processes that shape metallicity gradients vary with galaxy mass. The model identifies a transition at $\rm log(M_\star/M_{\odot})  \sim10^{10}-10^{10.5}M_{\odot}$, separating an advection-dominated regime at low masses from an accretion-dominated regime at high masses, where cosmic gas accretion introduces metal-poor gas that dilutes central regions and flattens the gradients in massive galaxies. 
Gradients are steepest at intermediate masses, where centrally-peaked metal production is the dominant process relative to both advection and accretion.


The absence of a clear turnover in $\log(\mathrm{N/O})$ gradients at $\log(M_{\star}/M_{\odot}) \sim 10.5$, further supports the above discussed picture of chemical equilibrium in the inner discs \citep{belfiore2017, Schaefer2020}.
The systematic steepening of the $\log(\mathrm{N/O})$ versus $12+\log(\mathrm{O/H})$ slope with stellar mass (see Figure~\ref{Figure:metallicity_NO_plot}) reflects the dual nucleosynthetic origin of nitrogen \citep{1990MNRAS.246..678E,2000ApJ...541..660H,2006MNRAS.372.1069M}: at low metallicities, nitrogen is produced as a primary element
and hence shows less dependence on $12 + \log(\mathrm{O/H})$ ratio, 
whereas at the high metallicities characteristic of massive galaxy centres, secondary CNO-cycle nitrogen 
production dominates, causing $\log(\mathrm{N/O})$ to rise steeply with O/H \citep{Alloin1979, Vincenzo2016}. 
The superlinear $\log(\mathrm{N/O})$ versus $12+\log(\mathrm{O/H})$ slope in the highest mass bin, therefore, reflects the radial metallicity gradient, placing the metal-rich centres in the secondary nitrogen regime while the metal-poor outskirts remain in the primary regime, naturally explaining why $\nabla\log(\mathrm{N/O})$ is systematically steeper than $\nabla\log(\mathrm{O/H})$ at high stellar masses \citep{belfiore2017, 
Poetrodjojo2018, 2021A&A...655A..58Z}. At low stellar masses, where both centres and outskirts remain in the primary nitrogen regime, the two gradients are comparable in magnitude, consistent with the near 1:1 relation seen in the insets (see Fig.~\ref{Figure:metallicity_NO_plot}).


At fixed stellar mass, the radial metallicity distribution does not show any morphological dependence. This may likely imply that the radial distribution of gas-phase metalallicity is primarily regulated by the depth of the gravitational potential well — traced by stellar mass — rather than by the internal structural configuration of the galaxy. Note that our sample is primarily designed for disc-dominated star-forming galaxies, and thus under-represents the bulge-dominated systems whose profiles remain poorly constrained.
%
%
A detailed morphological study based on full bulge-disk photometric decomposition, applied to a larger sample with equal representation of disk- and bulge-dominated systems across all stellar mass bins, would help to assess the role of galaxy structure in shaping radial abundance profiles.


The dependence of the metallicity profiles on galaxy size, at fixed mass and radii, suggests that physical compactness plays an important role in regulating the chemical enrichment. Since the two size sub-samples are matched in stellar mass and $D_{n}4000$, the observed offset cannot be attributed to differences in total stellar content or mean stellar age.
The broadly similar $D_{n}4000$ profiles, with a hint of slightly lower values in larger galaxies, suggest comparable recent star formation histories, with larger galaxies experiencing marginally higher recent star formation activity likely concentrated in their more extended discs. 
The associated accretion of metal-poor gas to fuel this star formation would dilute the local metallicity, particularly at large radii, naturally giving rise to the steeper negative gradients observed in extended galaxies compared to their compact counterparts of the same stellar mass and $D_{n}4000$.

This is in good agreement with \citet{Ellison2008}, who first reported a size--metallicity relation at fixed stellar mass, and with subsequent IFU studies ( \citealt{Boardman2021MNRAS} and \citealt{Lin2024}), confirming that compact galaxies exhibit higher metallicity and shallower gradients than extended galaxies across a wide range of stellar masses. 
The deeper gravitational potential wells of compact galaxies are more effective at retaining metal-enriched gas against supernova-driven outflows \citep{2004ApJ...613..898T, Finlator2008}, while under the inside-out growth paradigm, extended galaxies build up their outer discs through accretion of pristine gas, naturally steepening their gradients.

The systematic difference in metallicity gradients between galaxies with younger ($\nabla D_{n}4000 < 0$) and older ($\nabla D_{n}4000 > 0$) outskirts indicates that the age of the stellar population in the galactic disk plays a major role in shaping the metallicity gradients.
%
Galaxies with younger outskirts are actively building their discs inside-out, sustaining steep metallicity gradients through ongoing centrally concentrated star formation \citep{Prantzos2000, Pilkington2012}, while galaxies with older outskirts show near-flat gradients across all masses, consistent with more spatially uniform star formation histories. This is qualitatively consistent with \citet{2021A&A...655A..58Z}, who reported that galaxies with a positive $D_{n}4000$ radial gradient tend to have flatter O/H and N/O gradients, interpreting this as evidence for the evolution of metallicity gradients with stellar population age. 


\section{Summary and Conclusions}
\label{sec:conclusions}

In this work, we have exploited the unique targeting strategy and high-multiplexing spectroscopic capability of DESI to investigate the spatially resolved chemical enrichment in local star-forming galaxies. By utilizing multiple DESI fibers probing different regions within individual galaxies, we effectively use the survey as a sparse IFS to construct radial abundance profiles across a sample of 2291 star-forming galaxies, spanning $\sim$4 orders of stellar mass, and out to disc-halo interface. This represents one of the largest studies of gas-phase chemical abundance gradients in the local Universe. Splitting the sample into six stellar mass bins, and using empirical $T_{\mathrm{e}}$-based calibrations, we find:

\begin{enumerate}

\item The star-forming galaxies across all the stellar mass bins show a negative oxygen and nitrogen abundance gradient. Spanning from low- to high-mass galaxies, the gradients become progressively steeper, which is consistent with an inside-out growth scenario in which central regions undergo earlier and more efficient star formation and chemical enrichment, relative to the outer disc. Interestingly, the nearly flat radial profiles in the dwarf regime ($\log(M_\star/M_\odot) < 9.5$), suggest a more efficient redistribution of metals via feedback-driven outflows in shallower gravitational potentials.

\item We find a linear correlation between the metallicity gradient and stellar mass up to $\log(M_\star/M_\odot) \sim 10.5$, with gradients varying from 0.02 to 0.08 dex/R$_{e}$, beyond which the gradients become shallower - marking a clear turnover in the mass--metallicity gradient relation. This turnover is primarily driven by the flattening of metallicity profiles in the innermost regions ($\leq 1~R_{\mathrm{e}}$) of massive galaxies, where the central regions have reached chemical equilibrium following early and efficient star formation, stalling further enrichment, supported by the persistence of steep $\log(\mathrm{N/O})$ gradients. Cosmic gas accretion may additionally contribute by introducing metal-poor gas in the center that further shallows the gradients at high masses.


\item For the first time, probing the extended radial metallicity profiles out to $\sim5\,R_{\mathrm{e}}$, across different mass bins, we find that metallicity profiles flatten near the disk-halo interface. Most strikingly, the metallicities from different stellar mass bins converge in the outskirts. It supports the metallicity flooring predicted in cosmological simulations, which likely reflects reduced star formation efficiency in the outer disc, combined with the influence of metal-poor gas inflows that homogenize chemical abundances at large galactocentric distances.

\item Subdividing the sample by morphology (Sérsic index) reveals no significant difference in the metallicity or N/O gradients between disc-like and bulge-dominated galaxies. However, at fixed stellar mass and $D_{n}4000$, galaxy size is found to be a significant secondary parameter governing metallicity profiles, with compact galaxies retaining higher and more uniformly distributed metal content by resisting the metal-poor gas dilution associated with marginally higher star formation rate in more extended galaxies.

\item Splitting by their internal stellar age structure ($\nabla D_{n}4000$), galaxies with $\nabla D_{n}4000 < 0$ (younger outskirts) exhibit systematically steeper negative metallicity and $\log(\mathrm{N/O})$ gradients compared to galaxies with $\nabla D_{n}4000 \geq 0$ (older outskirts) at fixed stellar mass, 
suggesting that chemical enrichment gradients are further associated with radial stellar age structure.
The near-flat gradients in older-outskirt galaxies are consistent with more spatially uniform star formation histories, while the steeper gradients in younger-outskirt galaxies reflect ongoing inside-out disc growth.

\end{enumerate}

Together, they point towards a coherent picture of inside-out galaxy growth, where the interplay between star formation efficiency, stellar feedback, and metal-poor gas accretion collectively govern the radial chemical structure of galaxies across the galaxy population.


\begin{acknowledgments}
VN is supported by Beijing Natural Science Foundation (Grant No. IS25004). LCH was supported by the National Science Foundation of China (12233001), the National Key R\&D Program of China (2022YFF0503401), and the China Manned Space Program (CMS-CSST-2025-A09). AA acknowledges support from the INAF Large Grant 2022 “Extragalactic Surveys with JWST” (PI Pentericci) and from the European Union – NextGenerationEU RFF M4C2 1.1 PRIN 2022 project 2022ZSL4BL INSIGHT. VN acknowledges the Indian Institute of Astrophysics, Bengaluru for the hospitality during his visit and also thanks Haonan Zheng and Linhua Jiang for helpful correspondence and discussions.

\end{acknowledgments}





\clearpage
\bibliography{sample701}{}
\bibliographystyle{aasjournalv7}

\end{document}